\documentclass[aps,pop,nobibnotes,preprint,superscriptaddress,showkeys,amsmath,floatfix]{revtex4-1}  

\usepackage[colorlinks]{hyperref}

\usepackage{graphicx}
\usepackage{import}
\usepackage{xcolor}
\usepackage{textcomp}
\usepackage[capitalize]{cleveref}
\usepackage{placeins}
\usepackage[utf8]{inputenc}
\usepackage{xstring}
\usepackage{letltxmacro}

\usepackage{amssymb,amsfonts,amstext,graphics,graphicx,subfigure}
\usepackage{color}

\LetLtxMacro\origcite\cite
\newcommand{\citer}[1]{%
\begingroup
\def\tempx{0}%
  \StrCount{#1}{,}[\tempx]%
  \ifnum\tempx > 0 
  Refs. %
  \else
  Ref. %
  \fi
\endgroup
\origcite{#1}%
}

\begin{document}

\title{Fluid Flow Induced Deformation of a Boundary Hair}
\author{Jonas P Smucker\footnote[1]{Both authors contributed equally to this work}}
\email{jsmucker@chaos.utexas.edu}
\author{Zerrin M.\ Vural
\footnotemark[1]\footnote{present address:  Department of Mathematics UCLA.}}
\author{Jos\'e R. Alvarado}
\email{alv@chaos.utexas.edu}
\affiliation{Department of Physics and Center for Nonlinear Dynamics}
\author{Philip J.~Morrison}
\email{morrison@physics.utexas.edu}
\affiliation{Department of Physics and Institute for Fusion Studies, 
The University of Texas at Austin, Austin, TX, 78712, USA}


\begin{abstract}

{The deformation of a dense carpet of hair due to Stokes flow in a channel can be  described by a  nonlinear 
integro-differential equation for the shape of a single hair, which possesses  several solutions for a given choice of parameters. While being posed in a previous study and bearing resemblance to the pendulum problem from mechanics, this equation has not been analytically solved until now. Despite the presence on an integral with a nonlinear functional dependence on the dependent variable, the system is integrable. We compare the analytically obtained solution to a finite-difference numerical approach, identify the physically realizable solution branch, and  briefly study the solution structure through a conserved energy-like quantity. Time-dependent fluid-structure interactions are a rich and complex subject to investigate and we argue that the solution discussed herein can be used as a basis for understanding these systems.}
\end{abstract}

\maketitle


\section{Introduction}
\label{sec:intro}

Beds of hair-like structures {interacting with fluids} are prevalent in organisms on both micro and macro length scales. Their ubiquity in complex and simple organisms is an indication of their versatility. Indeed, there is great diversity in the functionality at either of these length-scales.

For example, geckos utilize hair-like setae on their feet to promote adhesion to surfaces \cite{autumn_evidence_2002}, cricket filiform hairs play a mechanosensitive role \cite{cummins_interaction_2007}, and the papillae on hummingbird tongues are used as a ``nectar mop" \cite{harper_specialized_2013}. They serve important roles in nutrient absorption \cite{reicher_intestinal_2021,zou_relationships_2019}, surface protection and flow control \cite{luhar_flow-induced_2011,weinbaum_glycocalyx_2021,chateau_why_2019,angleys_kroghs_2020}, surface adhesion \cite{autumn_evidence_2002,walker_adhesive_1985,bullock_beetle_2011,suter_taxonomic_2004}, and fluid entrainment \cite{harper_specialized_2013,kim_optimal_2011,nasto_viscous_2018}. They function as mechanosensors, detecting fluid flows \cite{weinbaum_glycocalyx_2021,cummins_interaction_2007,guo_hydrodynamic_2000,hood_marine_2019,thomazo_probing_2019,thomazo_collective_2020,takagi_active_2020}, predators \cite{Chagnaud4479,10.1242/jeb.02485}, and electric fields \cite{sutton_mechanosensory_2016}.

With the improvement of existing manufacturing techniques and the creation of new protocols \cite{du_roure_dynamics_2019,nasto_air_2016,hanasoge_asymmetric_2017,zhang_tailoring_2021,wang_continuous_2016,paek_microsphere-assisted_2014}, studies have investigated increasing small, high aspect-ratio systems of artificial hairs. For example, recent studies have  investigated  the design potential of hair beds:   Hairs placed in a microfluidic channel have been shown to function as pumps \cite{wang_continuous_2016,zhang_metachronal_2021,milana_metachronal_nodate}, rectifiers \cite{alvarado_nonlinear_2017,stein_coarse_2019}, and micro-mixers \cite{shanko_microfluidic_2019,zhang_transport_2021,saberi_stirring_2019} making them a design consideration in lab-on-chip devices.

Earlier work  \cite{alvarado_nonlinear_2017} used the theory of Kirchoff rods to describe the bending of hairs in a channel when subject to shear flow. These authors assumed that the hairs possess linear, isotropic material properties, but undergo finite displacements. The latter consideration makes the problem nonlinear \cite{audoly_elasticity_2010} and,  as a result, in \cite{alvarado_nonlinear_2017}  the problem was solved numerically.

{
While there are many numerical methods to deal with such nonlinearities, numerical approaches will only go so far. Biological-scale simulation of hair-beds has yet to be achieved efficiently \cite{luminari_modeling_nodate}. There are several reasons for this. Such systems involve many hairs \cite{stein_coarse_2019} that are free to respond to the ambient fluid flows generated by both external forcing and their neighbors. Additionally, consideration of the hair's inertia makes the governing system of equations stiff \cite{audoly_elasticity_2010}. Despite this, large-scale simulation of hairs has been achieved in the graphics community by application of an assortment of optimization techniques \cite{petrovic_volumetric_2006,ryu_500_2007}. However, these techniques have the trade-off of realism \cite{iben_artistic_2013}.
}

To further understand these systems, we focus on and solve just the time-independent problem  posed in \cite{alvarado_nonlinear_2017} for the profiles of a cantilevered hair-bed subject to shear flow through a channel. We investigate both physical and nonphysical classes of solutions and  how to consistently single out the  former from the latter.

{The paper is organized as follows.} {In \cref{sec:formulation}, we introduce the basic model and examine how our problem differs from previous studies. We see that the problem arises naturally as a boundary value problem, for which a method for analytical solution is described and implemented in  \cref{sec:solution}.  Next, we examine the phase space and discuss how a self-consistency condition associated with the problem influences the solution-structure in \cref{sec:selfcon}. It is here we also consider the case of angled hairs. We discuss common numerical approaches to solving this class of problem in \cref{sec:hairProfiles}, comparing one such implementation to our solution. We conclude our work with a summary in \cref{sec:conclusion}.}


\section{Problem Formulation}
\label{sec:formulation}

\begin{figure}
    \centering
    \includegraphics[scale=1.5]{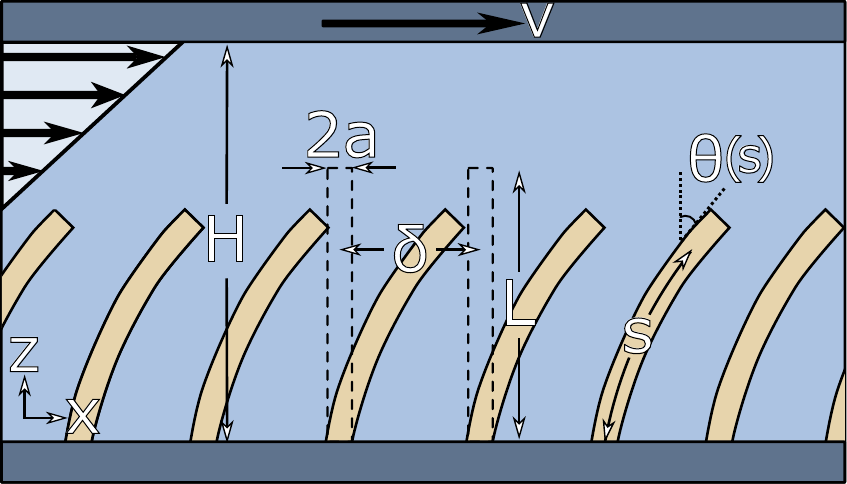}
    \caption{Illustration of how the fluid-hair system is modeled. Dashed and solid profiles show hairs in an undeformed and a deformed configuration, respectively. The model assumes the fluid velocity becomes zero at the hair-tip, exerting a shear stress $\frac{\eta v}{H-h(L)}$ on the hair.}
   \label{fig:model}
\end{figure}

The problem of a cantilevered hair, attached at a flat horizontal boundary, subject to Stokes flow is described by the following equation:
\begin{equation}
\label{TheEqn2}
EI\frac{d^2\theta(s)}{ds^2}=-\frac{\pi a^2}{\phi}\frac{\eta v \cos\theta(s)}{H-\int_0^L \cos\theta(s')\, ds'}\,,
\end{equation} 
where $\theta$ is the angle a tangent to the backbone of the hair makes with the vertical and $s$ is a parameter that measures the arc length along the hair,  taken to range from $0$ to $L$. As is evident from \eqref{TheEqn2}, the problem has several parameters,  which we   summarize  in  \cref{fig:model,tab:vars}.  These include four length scales: the diameter of the hair $a$, the length of the hair $L$,  the height of the channel $H$, and the {hair to hair centerline spacing, $\delta$. Instead of using $\delta$ explicitly, we use the dimensionless packing fraction $\phi=\frac{2\pi}{\sqrt{3}}\frac{a^2}{\delta^2}$ which quantifies how closely packed the hairs are.} In addition we have the hair's elastic modulus $E$, its second moment $I=\frac{\pi a^4}{4}$, the fluid viscosity $\eta$, and the imposed fluid velocity $v$. 

\begin{table}
\centering
    \caption{Important system parameters and their associated units\label{tab:vars} }
    \begin{tabular}{ll}
        \hline
        $H$&Channel height $[L]$\\
       $L$& Hair length $[L]$\\
         $a$&Hair radius $[L]$\\
            $\phi$&{Packing fraction}\\
        {$\delta$}&{Hair-hair centerline spacing $[L]$}\\
         $E$&Elastic modulus $[M][T]^{-2} [L]^{-1}$  \\
         $I$& 2nd area moment of hair's cross-section $[L]^4$\\
         $\eta$& {Dynamic} viscosity $[M][L]^{-1}[T]^{-1}$\\
         $v$&{Imposed} fluid velocity $[L][T]^{-1}$\\
         $\theta$&Angle between the local tangent and the vertical\\
         $s$&Arc length measured from the base of the hair $[L]$\\
    \end{tabular}
\end{table}

In this formulation the hair is represented as a plane curve in Cartesian coordinates with $\mathbf{R}=x(s) \hat{x} + z(s) \hat{z}$, where  $\hat{x}$ and  $\hat{z}$ are unit vectors.  The unit tangent is given by $\hat{\mathbf{T}}=\sin\theta(s)\,  \hat{x} + \cos\theta(s) \,\hat{z}$, so that  $\hat{\mathbf{T}}\cdot \hat{z}= \cos \theta$, and with this parameterization the curvature is given by $d\theta/ds$.  The quantity $h(s) = \int_0^s \cos\theta(s')\, ds'$ represents the height of the hair at position $s$ with $h(L)$ being the total height. Because the flow is horizontal (in the $\hat{x}$-direction), the force on the hair depends on its vertical height, and this is the reason for the denominator ${H-\int_0^L \cos\theta(s')\, ds'}= H-h(L)$ of \eqref{TheEqn2}: a vertical hair would be  impacted by a maximum force,  while the force is diminished as it entrains in the horizontal direction.  A most interesting feature of this formulation is that the relaxed state that results is self-referential because of this denominator; i.e.,   the solution depends on itself and, as we will see,  this gives rise to a self-consistency condition.

{
The form of \eqref{TheEqn2} is obtained from moment balance in equilibrium. For an infinitesimal cylindrical section of a rod (the hair) this balance yields
\begin{equation}
\label{eq:momentBalance}
\mathbf{M}(s+ds)-\mathbf{M}(s) + \mathrm{d}\mathbf{r}\times \mathbf{F}_{int}(s)=0\, ,
\end{equation}
where $\mathbf{M}(s)=EI\theta'(s) \hat{\mathbf{y}}$ is the bending moment, $\mathrm{d}\mathbf{r}=\mathrm{d}s\hat{\mathbf{T}}(s)$, and $\mathbf{F_{int}}(s)$ is the net internal force on the rod segment. By ``dividing" by $\mathrm{d}s$, we obtain \eqref{TheEqn2}.
}
We refer the reader to   \cite{audoly_elasticity_2010,alvarado_nonlinear_2017} for further details.

 Equation \eqref{TheEqn2} can be transformed into  the compact nondimensional form  
 \begin{equation}\label{hairdiffeq}
    \frac{d^2\hat\theta}{d\sigma^2}=-\omega_\epsilon^2\cos\hat\theta,
\end{equation}
by introducing 
\begin{equation}
\hat\theta(\sigma) = \theta(s)\,,\quad \sigma= \frac{s}{L}\,,\quad \epsilon=\frac{L}{H}\,, \quad\mathrm{and}\quad  \omega^2=\frac{\pi a^2 L^2\eta v}{EIH\phi}\,,
\end{equation}
and 
\begin{equation}
\label{oe-givens}
    \omega_\epsilon^2=\frac{\omega^2}{1-\epsilon\int_{0}^{1}\cos\hat{\theta}(\sigma)\,d\sigma}\,.
\end{equation}
The natural boundary conditions for \eqref{hairdiffeq}  are the following: 
\begin{equation}
\label{eq:bcs}
    \hat\theta(0)=\hat\theta_0 \quad\mathrm{and}\quad    \left.\frac{d\hat\theta}{d\sigma} \, \right|_{\sigma=1} \!\!\! \!= \, \hat\theta'_1=0\,,
\end{equation}
where $\hat\theta_0$ is the angle of attachment of the hair and ${\hat\theta}_1'=0$ means that the hair at its tip has zero curvature. {The latter condition can be obtained from the moment balance in \cref{eq:momentBalance}. At the end of the hair, there is no upstream ($s>L$) contribution to the balance implying that $\hat{\mathbf{M}}(1)\equiv EI\hat{\theta}_1'\hat{\mathbf{y}}$ is infinitesimally small.} With these definitions, we see that our system has only two dimensionless parameters, $\epsilon$ and $\omega$, in addition to the choice of  $\hat\theta_0$. We will drop the `hats' moving forward to avoid clutter.

This system differs in  some  ways from the standard pendulum problem of mechanics.  For example, we have the trivial difference that there is a shift in the definition of the angle -- instead of having   $\sin\theta$ on the righthand side of \eqref{hairdiffeq} we have $\cos\theta$.  However, there are two essential differences: first, instead of the usual  initial value problem, in light of \eqref{eq:bcs}, we have a boundary value problem and second, the system has the self-referential feature mentioned above, i.e.,  in order to know the effective frequency $\omega_{\epsilon}$ of \eqref{oe-givens} one must first obtain the entire ``orbit" $\theta(\sigma)$ to get a self-consistent solution.  Although these differences significantly complicate the problem,  we will see that the problem remains integrable.   We will see that the self-consistency condition together with the  boundary value nature of the problem  lead to a sort of quantization and a further reduction of parameters.


%
%
%
%
%
%

A ``potential" for \eqref{hairdiffeq} can be obtained by
 setting $d^2\theta/d\sigma^2=-dV/d\theta$, where $V(\theta)=\omega^2_\epsilon\sin\theta$. The Hamiltonian of this system, which we will call $\mathcal{E}$,  takes the form
\begin{equation}
\label{hairH}
    \mathcal{E}=\frac{1}{2}\left(\frac{d\theta}{d\sigma}\right)^2+\omega^2_\epsilon\sin\theta\,.
\end{equation}
Because $\omega_\epsilon$ does not depend explicitly on $\sigma$, the time-like variable, conservation of energy, $d     \mathcal{E}/d\sigma=0$, follows immediately.  
Observe that the curvature, $d\theta/d\sigma$,  determines a quantity analogous to the pendulum kinetic energy for this system.  

 As noted above, in the pendulum problem the potential is $-\cos\theta$ and the pendulum oscillates about $\theta = 0$. However, the boundary value problem for the hair is  different because the pendulum potential $-\cos\theta$ is shifted by $\pi/2$ from the hair's potential, $\sin\theta$. Thus,  the hair problem is analogous to a pendulum starting at $\theta=\theta_0$, a distance up the potential well, that is  then projected  further up the well with an initial velocity that is enough for it to hit its turning point at $d\theta/d\sigma = 0$. Therefore, the goal is  to determine the initial value of $d\theta/d\sigma$ corresponding to a time (length) for this to occur. To transform our problem to the pendulum problem we will shift $\theta$ by $\pi/2$, i.e., 
\begin{equation}
 \bar{\theta}=\theta+\frac{\pi}{2}\qquad \Rightarrow \qquad \sin(\theta)=-\cos(\bar{\theta}) \,,
\end{equation}
and therefore
\begin{equation}
\label{hairHbar}
    \mathcal{E}=\frac{1}{2}\left(\frac{d\bar\theta}{d\sigma}\right)^2\!- \, \omega^2_\epsilon\sin\bar\theta\,.
\end{equation}
 This angle shift is convenient because it allows us to write the solution in the standard form for the pendulum in terms of elliptic integrals, which is a first step toward showing integrability.

%


\section{Solution-Integrability}
\label{sec:solution}

Given the formulation of \cref{sec:formulation}, we may begin by following the elementary  procedure for reducing the pendulum to quadrature.  Using the double-angle formula, $\cos\theta=1-2\sin^2(\theta/2)$, solving  \eqref{hairHbar} for $d\bar{\theta}/d\sigma$,  and integrating gives
\begin{equation}
\label{eq:quadrature}
     {\pm}\frac{\omega_{\epsilon}}{k}\, \sigma= \int^{\theta(\sigma)/2+\pi/4}_{{\theta_0}/{2}+ \pi/4} \frac{d{\chi}}{\sqrt{1-k^2\sin^2\chi}}\,, 
\end{equation}
where $\chi=\theta/2$ and 
\begin{equation}
\label{k2}
k^2 = \frac{2\omega_\epsilon^2}{\mathcal{E}+ \omega_\epsilon^2} \,. 
\end{equation}
{The choice in sign in \eqref{eq:quadrature} determines whether $\theta_0'$ is positive or negative. While we are primarily interested in hairs with positive base-curvature (corresponding to the positive sign), we include both possibilities for completeness.}   This quadrature, analogous to that of the pendulum, is the first step toward obtaining integrability of our hair problem. 

Before proceeding, there is one issue that must be checked, viz.\  that $\sqrt{1-k^2\sin^2\chi}$ does not become imaginary; that is, we want to check that  $k^2\sin^2\chi <1$ for $\chi$ within the limits of integration, and that  this is maintained as the  upper limit of the integral of \eqref{eq:quadrature} extends all the way to $\theta(1)$, which we will denote by $\theta_1$.  For the most part, we expect physical solutions to have 
\begin{equation}
\label{th1rng}
0\leq\theta_1 \leq \pi/2 \,,
\end{equation}
which we can  verify after the solution is obtained, so that according to \eqref{hairH}, $\mathcal{E} >0$.  Thus, upon  writing $\xi=\mathcal{E}/\omega^2_{\epsilon}$,  \eqref{k2} becomes $k^2= 2/(1+\xi)$ with $\xi\geq 0$. Consequently,
\begin{equation}
 1\leq k^2 \leq 2\,,
 \label{krange}
 \end{equation}
  and this by itself is insufficient to guarantee  $k^2\sin^2\chi <1$, However, because of  the second boundary condition of \eqref{eq:bcs} and conservation of the energy of \eqref{hairHbar}
\begin{equation}
\label{xi}
    \xi=  \sin\theta_1\quad \Rightarrow \quad 0 \leq \xi\leq 1 \,.
\end{equation}
Next, using  \eqref{th1rng}  and the fact that $\sin^2\chi$ achieves its maximum when $\theta=\theta_1$, we obtain
\begin{equation}
\sin^2\big(\theta_1/2 + \pi/4\big)=\big( 1 + \sin\theta_1\big)/2\,,
\end{equation}
which follows from elementary trigonometry identities.  Therefore with \eqref{xi}, we have
\begin{equation}
k^2\sin\chi\leq \frac{2}{1 + \xi}\,  \frac12\, \big( 1 + \sin\theta_1\big)= \frac{1}{1 + \xi} \, \big( 1 + \xi \big)=1\,.
\end{equation}
Thus the quadrature integral of \eqref{eq:quadrature} is well behaved even with $k^2>1$, which is consistent with what  we would physically expect. 

Proceeding, we can invert and obtain the explicit solution by writing  the integral of  \eqref{eq:quadrature} in terms of elliptic integrals. First, we split the integral as follows:
\begin{equation}
\label{split}
   {\pm}\frac{\omega_\epsilon}{k}\,\sigma =\hphantom{+} \int^{\theta(\sigma)/2+\pi/4}_{0}\hspace{-.8cm}\frac{d\chi}{\sqrt{1-k^2\sin^2\chi}} 
    - \int_0^{\theta_0/2 + \pi/4}\hspace{-.8cm}\frac{d\chi}{\sqrt{1-k^2\sin^2\chi}}\,,
\end{equation}
and notice that the second integral of \eqref{split} is an incomplete elliptic integral of the first kind, which we move to the lefthand side,  yielding 
\begin{equation}
\label{reduce}
      {\pm}\frac{\omega_\epsilon}{k}\,\sigma  + F\left(\left.\frac{\theta_0}{2} + \frac{\pi}{4}\,  \right|  k^2\right) =F\left(\left.\frac{\theta(\sigma)}{2}+\frac{\pi}{4}\,\right| k^2\right) \,.
\end{equation}

Equation \eqref{reduce} can be inverted by utilizing Jacobi elliptic functions. In particular, the Jacobi amplitude function (see e.g.\ \cite{abramowitz_handbook_1965}) is the inverse of  $F$ , i.e., 
\begin{equation}
\label{amfun}
    \mathrm{am}\big(F(\phi\vert k^2)\vert k^2\big)=\phi \,.
\end{equation}
From now on we will drop the $k^2$ from the arguments and write  $\mathrm{am}(\phi)$ for $\mathrm{am}(\phi\, \vert\,  k^2)$ and  $F(\phi)$ for $F(\phi\vert k^2)$, unless a different parameter is used.
Using \eqref{amfun}, \eqref{reduce} can be inverted  to obtain the following solution:
\begin{equation}
\label{nonSol}
    \theta(\sigma)=2\,  \mathrm{am}\left({\pm}\frac{\omega_\epsilon}{k}\,\sigma  +F\left(\frac{\pi}{4}+\frac{\theta_0}{2}\right)\right)-\frac{\pi}{2}.
\end{equation}
Evaluation of  the Hamiltonian of \eqref{hairHbar} at $\sigma=0$ gives
\begin{equation}
\label{E0}
\mathcal{E}= \frac12 \left({\theta_0'}\right)^2 +  \omega_\epsilon^2\,\sin\theta_0\,,
\end{equation}
 where  $\theta_0'= d\theta(0)/d \sigma$.  Using  \eqref{k2} and \eqref{E0} we see that \eqref{nonSol} gives 
$\theta(\sigma, \theta_0, \theta_0',\omega_\epsilon)$, as expected for the solution of the initial value problem. 
To solve the boundary value problem where $\theta_1'=0$ we use the identity ${d\, \mathrm{am}(u)}/{du}= \mathrm{dn}(u)$ and hence, 
\begin{equation}
\label{thprime}
\frac{d\theta(\sigma)}{d\sigma}={\pm}2\, \frac{\omega_\epsilon}{k}\, \mathrm{dn}
\left({\pm}\frac{\omega_\epsilon}{k}\,\sigma  +  F\left( \frac{\pi}{4}+ \frac{\theta_0}{2}\right) \right),
\end{equation}
and therefore the boundary condition gives 
\begin{equation}
\label{BC}
\theta_1'={\pm}2\,\frac{\omega_\epsilon}{k}\, \mathrm{dn}
\left({\pm}\frac{\omega_\epsilon}{k}  +  F\left(\frac{\pi}{4}+ \frac{\theta_0}{2}\right) \right)=0\,.
\end{equation}
Because elliptic integrals and functions usually consider the range $0\leq k^2\leq 1$, while we have \eqref{krange},  we use the identity
\begin{equation}
\mathrm{dn} (u\,|\,k^2)= \mathrm{cn}(ku\,  |\, k^{-2})
\end{equation}
to write the  boundary condition of \eqref{BC} in the form
\begin{equation}
 \mathrm{cn}\left({{\pm}\omega_\epsilon}  +  k F\left( \left.\frac{\pi}{4}+ \frac{\theta_0}{2}\, \right|\, k^2\right)\Bigg| k^{-2} \right)=0\,.
 \label{condition}
\end{equation}
Equation \eqref{condition}  gives a condition relating $\theta'_0$ to ${\omega_\epsilon}$ for fixed $\theta_0$.  Because of the periodic nature of cn$(u)$,  these are quantized according to 
\begin{equation}
{\pm}{\omega_\epsilon}  +  k F\left( \left.{\pi}/{4}+ {\theta_0}/{2}\, \right|\, k^2\right)=  {(2n+1)} K(k^{-2})\,,\qquad n\in\mathbb{Z}\,,
\end{equation}
where $K(k^{-2})= F(\pi/2\, |\,k^{-2})$.

To summarize we collect all our parameters together, 
\[
k^2= \frac{2}{1+\xi}\,, \qquad \xi=\frac{\mathcal{E}}{\omega^2_\epsilon}\,,
\qquad  \mathcal{E}=\frac12 (\theta'_0)^2 + {\omega_\epsilon}^2\sin\theta_0\,, 
\]
and observe, we have shown  for fixed and given ${\omega_\epsilon}$ and $\theta_0$ the above analysis  tells us what $\theta'_0$ must be to hit our boundary condition $\theta_1'=0$. 

So far we have followed a conventional and straightforward path leading to the solution of \eqref{nonSol}. Except for the shift in phase and the boundary value nature of this solution, it is standard for a one degree-of-freedom Hamiltonian system: it depends on two parameters related to possible initial conditions $\theta_0$ and $\theta_0'$ via $\mathcal{E}$ and one parameter $\omega_{\epsilon}$,  which we have  treated as a given constant.  We proceed now by examining in general terms the boundary value nature of our problem with  the imposition of  the self-consistency constraint of \eqref{oe-givens}. 

Consider a general system of differential equations of the form
\begin{equation}
\frac{d^2\theta}{d\sigma^2}= f(\theta, \lambda)\,,
\label{general}
\end{equation} 
where $\lambda$ is a parameter.  Often one uses a shooting method to solve the boundary value problems for equations of this type, i.e., a sequence of initial conditions are integrated numerically for choices of  the parameter $\lambda$ until the desired  boundary condition is reached.  This procedure usually selects out discrete values for $\lambda$,  which for linear systems would be eigenvalues.  However, if one has an analytical solution to the initial value problem, as we do, this can be used to relate initial and final values. A condition that relates derivatives at the endpoints, here taken to be $\sigma=0$ and $\sigma=1$, follows immediately upon integrating \eqref{general}, i.e.
\begin{equation}
\theta_1'-\theta_0'= \int_0^1 f(\theta, \lambda)\, d\sigma\,.
\label{sampcon}
\end{equation}
Self-consistency means that the  parameter  $\lambda$ depends functionally on the solution $\theta(\sigma)$.  For our problem at hand, the role played by $\lambda$ is $\omega_\epsilon$ and  this self-consistency  requires the  solution of \eqref{nonSol} be consistent  with the $\omega_\epsilon$ as calculated from \eqref{oe-givens} with the insertion of \eqref{nonSol}.    As a first step toward imposing this self-consistency  constraint, analogous to \eqref{sampcon} we integrate \eqref{hairdiffeq} to obtain an expression  for the height of the hair in terms of an initial condition, viz., 
\begin{equation}
\label{intOrig}
    \theta_0'=\omega_\epsilon^2\int_0^1\cos\theta(\sigma)\, d\sigma= \omega^2 \frac{h_1}{1-\epsilon h_1}=\frac1\epsilon\,(\omega_\epsilon^2-\omega^2) 
    \,,
\end{equation}
where  $\theta_0'= d\theta(0)/d \sigma$ and $h_1$ is the dimensionless height of the hair,  the dimensional height being  $h_1L$.  The last equality of \eqref{intOrig} follows upon eliminating $h_1$ using \eqref{oe-givens}.
The hair problem is complicated because the quantity $\omega^2_{\epsilon}$   depends on the solution of the boundary value problem \eqref{oe-givens}   to  give  \eqref{intOrig}.   Fortuitously, this quantity only depends on $h_1$, i.e., $\theta_0'$ is proportional to $h_1$ and  the constant of proportionality $ \omega^2_{\epsilon}=\omega^2/({1-\epsilon h_1})$ also depends on  $h_1$.  For general problems of this nature of the form of \eqref{sampcon}, these two quantities would not in general depend on a single parameter like this. 

Evidently, we must calculate $h_1$.  In fact, we can explicitly calculate $h(\sigma)$ the height of the hair at parameter value $\sigma$ (see Appendix \ref{appendix}), 
\begin{eqnarray}
h(\sigma)&=&\int_0^\sigma\! \cos\theta(\sigma')\, d\sigma'
\nonumber\\
&=& \frac{2}{k \omega_{\epsilon}}\left[ 
\sqrt{1 - k^2\sin^2 ({{\pi}/{4}+ {\theta_0}/{2}})} - \mathrm{dn}\left({\pm}\frac{\omega_{\epsilon}}{k}\, \sigma +  F\left( {\pi}/{4}+ {\theta_0}/{2}\right) \right)
\right].
\label{hs}
\end{eqnarray}
Next, we write $\omega_\epsilon$ in terms of $\omega$ and $\epsilon$ by inserting the last equality of  \eqref{intOrig} into \eqref{E0}, giving
\begin{equation}
    \omega_\epsilon^2\,{\mp}\,\epsilon\sqrt{2\mathcal{E}-2\omega_\epsilon^2\sin\theta_0}=\omega^2\,.
\end{equation}
Thus the self-consistent solution of our boundary value problem is fully determined by following:
\begin{equation}
\label{solution}
        \theta(\sigma;\theta_0,\epsilon,\omega)=2\,\mathrm{am}\left( {\pm}\omega_\epsilon\sqrt{\frac{\xi+1}{2}} \sigma
        +F\left(\frac{\pi}{4}+\frac{\theta_0}{2}\Big\vert\,k^2\right)\bigg\vert\, k^2 \right)-\frac{\pi}{2}\,,
 \end{equation}
where $0\leq \sigma \leq 1$ is our dimensionless parameter and 
\begin{eqnarray}
        0&=& {\pm} {{\omega_\epsilon}  +  k F\left( \left.{\pi}/{4}+ {\theta_0}/{2}\, \right|\, k^2\right)-  (2n+1)K(k^{-2})}\,,
        \label{solution2}\\
        \omega_\epsilon^2& =& \omega^2\,{\pm}\, \epsilon\omega_\epsilon\sqrt{2\xi-2\sin\theta_0}\,,
        \label{solution3}\\
        k^2&=&\frac{2}{\xi+1}
        \label{solution4}\,.
 \end{eqnarray}
Note, $kF(\varphi|k^2)=F(\bar{\varphi}|k^{-2})$, where  $\sin \bar\varphi= k \sin\varphi$, (see equation (8.127) of \cite{gradshteyn2007})  can be used when evaluating \eqref{solution2}.  Here \eqref{solution2} with  \eqref{solution4} determines $\omega_\epsilon$ as a function of $\theta_0$ and $\xi$, which  with \eqref{solution3} determines $\xi$ as a function of $\theta_0$, $\epsilon$,  and $\omega$.  We note in passing that the variable $\xi$ is related to  the physically perspicuous variable $h_1$ according to
 $$
 \xi
 =  \frac{\omega^2}{2} \frac{h_1}{\left( 1- \epsilon h_1\right)^2} +  \sin\theta_0 \,.
 $$
 
 In \cref{sec:selfcon} we will evaluate \eqref{solution} for various cases.  We will see that for    physically realizable solutions of interest, we must set $n=0$ in \eqref{solution2} and select the $+$ branch.  In practice we use root finding to solve \eqref{solution2} and \eqref{solution3}.   


\section{Phase Space Interpretation}
\label{sec:selfcon}

Because \eqref{hairdiffeq} is isomorphic to the differential equation for a pendulum, it is helpful to interpret our analytical solutions in terms of motion in the pendulum  phase space.  In this section we do this, first for hairs with $\theta_0=0$ and then for $\theta_0\neq 0$. 

\subsection{Vertical hairs: \texorpdfstring{$\theta_0=0$}{t0=0}}

\begin{figure}
    \centering
    \includegraphics[scale=1.75]{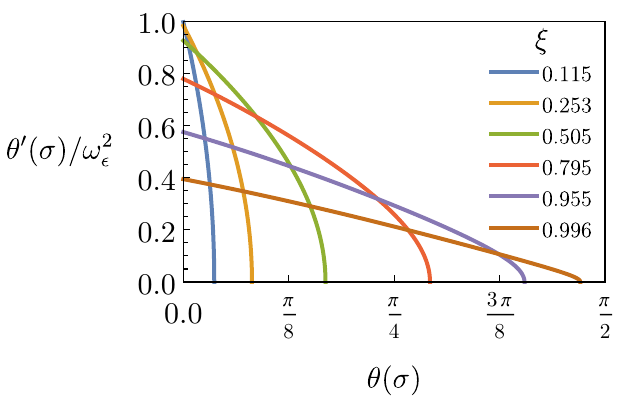}
    \caption{Phase portraits for a selected set of $0\leq\xi\leq1$, a scaled measure of the energy,  and their corresponding profiles. $\xi=1$ corresponds with the separatrix. Trajectories start at $\theta(0)=0$ and end at $\theta'(1)=0$.}
    \label{fig:phasePortrait}
\end{figure}

 \Cref{fig:phasePortrait} shows several different trajectories, corresponding to different values of $\xi$, for the case where $\theta_0=0$. Here, only the solutions that stop when they intersect $\theta_1'=0$ once are shown, but we do observe other solutions corresponding to trajectories completing one or several orbits, especially at higher values of $\xi$.

{Observe, $\xi={\mathcal{E}}/{\omega_\epsilon^2}$ and $\mathcal{E}$ are both measures of the system's energy (Hamiltonian) since they only differ by a proportionality constant,  once self-consistency is enforced.    We prefer to use $\xi$ in the following figures and analysis because $-1\leq \xi\leq1$,  while $\mathcal{E}$ is unbounded.  In addition, our analytic solution is written more concisely in terms of $\xi$.  \Cref{fig:phasePortrait} shows the phase space with energy surfaces parameterized by $\xi$.   Note, because $\xi$ is used and because the ordinate is $\theta'/\omega_\epsilon^2$,  the energy surfaces are not nested as usual.    In \cref{fig:phasePortrait}, as $\xi\rightarrow 1$ the orbit approaches the separatrix (within the pendulum analogy, a value of $\xi=1$ corresponds to a pendulum ``kicked" up from $\theta_0=0$ to $\theta_1=\frac{\pi}{2}$) and $\xi=0$ corresponds to the undeformed hair where $\theta(\sigma)\equiv 0$.}

Within the pendulum analogy, a fixed choice of  $\omega_\epsilon$ is related to a choice of gravitational acceleration. The boundary conditions $\theta_0=0$ and $\theta'_1=0$ describe a pendulum trajectory starting at $\theta=0$ and ending when $\dot{\theta}=0$ in a time $T$ (analogous to the length of a hair). The largest possible initial velocity (or energy) that satisfies these conditions corresponds to a phase-space trajectory entirely confined to the first quadrant. At a threshold, other starting velocities can also satisfy the ``initial" conditions, but they must correspond to orbits that exit the first quadrant.

\begin{figure}
    \centering
    \includegraphics[scale=1.25]{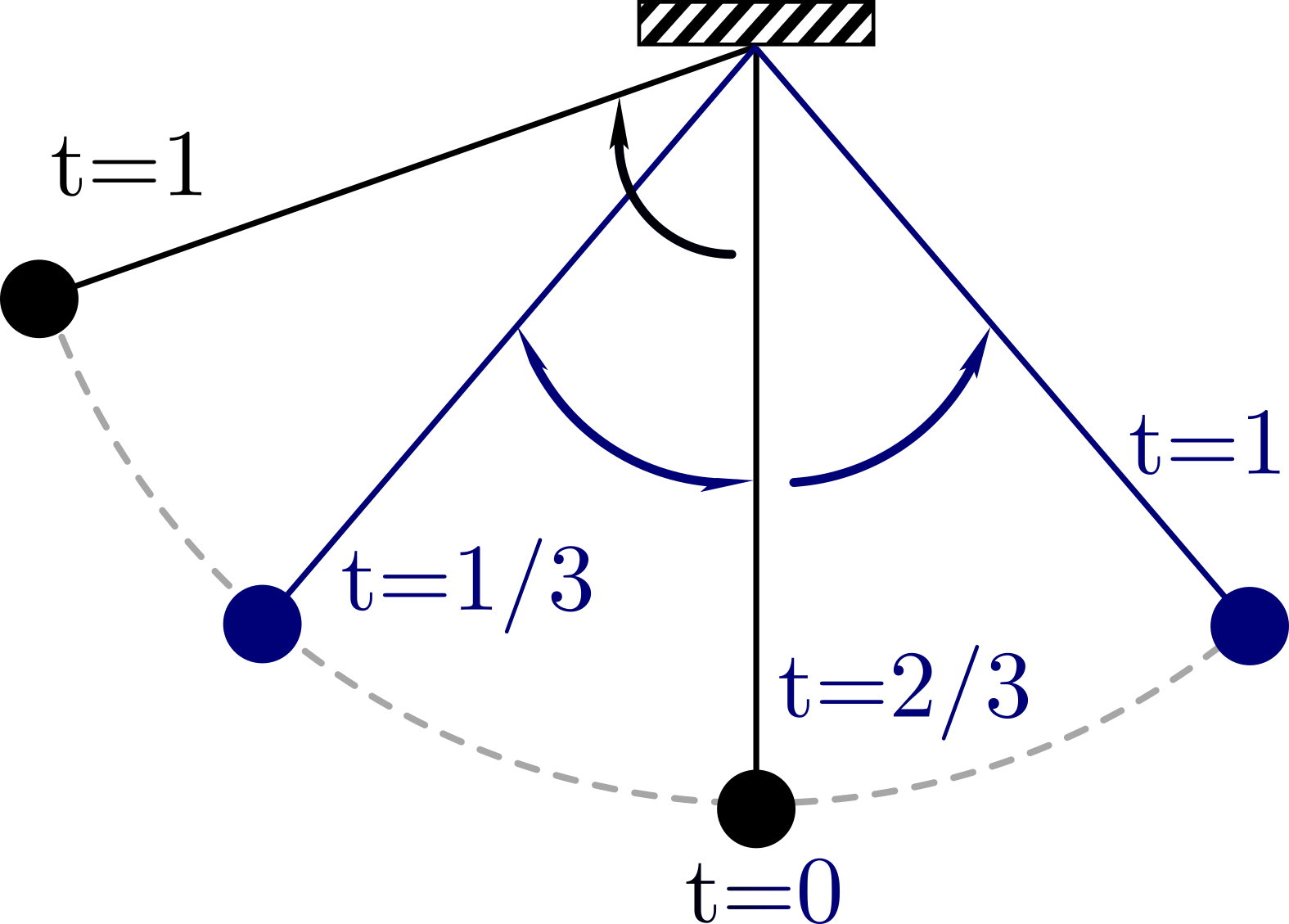}
     \caption{The problem of a cantilevered hair with a point load at its end is isomorphic to the equation of motion for a pendulum with the initial conditions $\theta(t=0)=0$ and $\dot{\theta}(t=1)$=0. Because multiple different orbits can satisfy these conditions for a given choice in parameters (e.g., the blue and black orbits of the figure), both the cantilevered hair and pendulum problem posed above are not unique.}
    \label{fig:pendulum}
\end{figure}

\Cref{fig:pendulum} depicts two orbits for a given choice of  parameters. The first (black) starts at $\theta_0=0$ with some $\dot{\theta}_0\neq 0$ and the trajectory evolves until $\dot{\theta}_1=0$. On the other hand, the blue orbit reaches its first maximum when $\dot{\theta}(t=1/3)=0$ and it oscillates the other direction until finally reaching $\dot{\theta}_1=0$. When not equal to zero, the branch index, $n$ (shown in \cref{solution}) selects out these lower period orbits.  {In addition to these two solutions in our example above, there are two more with an opposite sign in $\dot{\theta}_0$. This choice in direction is reflected by the $\pm$ sign in our solution.}

\begin{figure*}
    \centering
    \label{fig:solution}
    \includegraphics[scale=1]{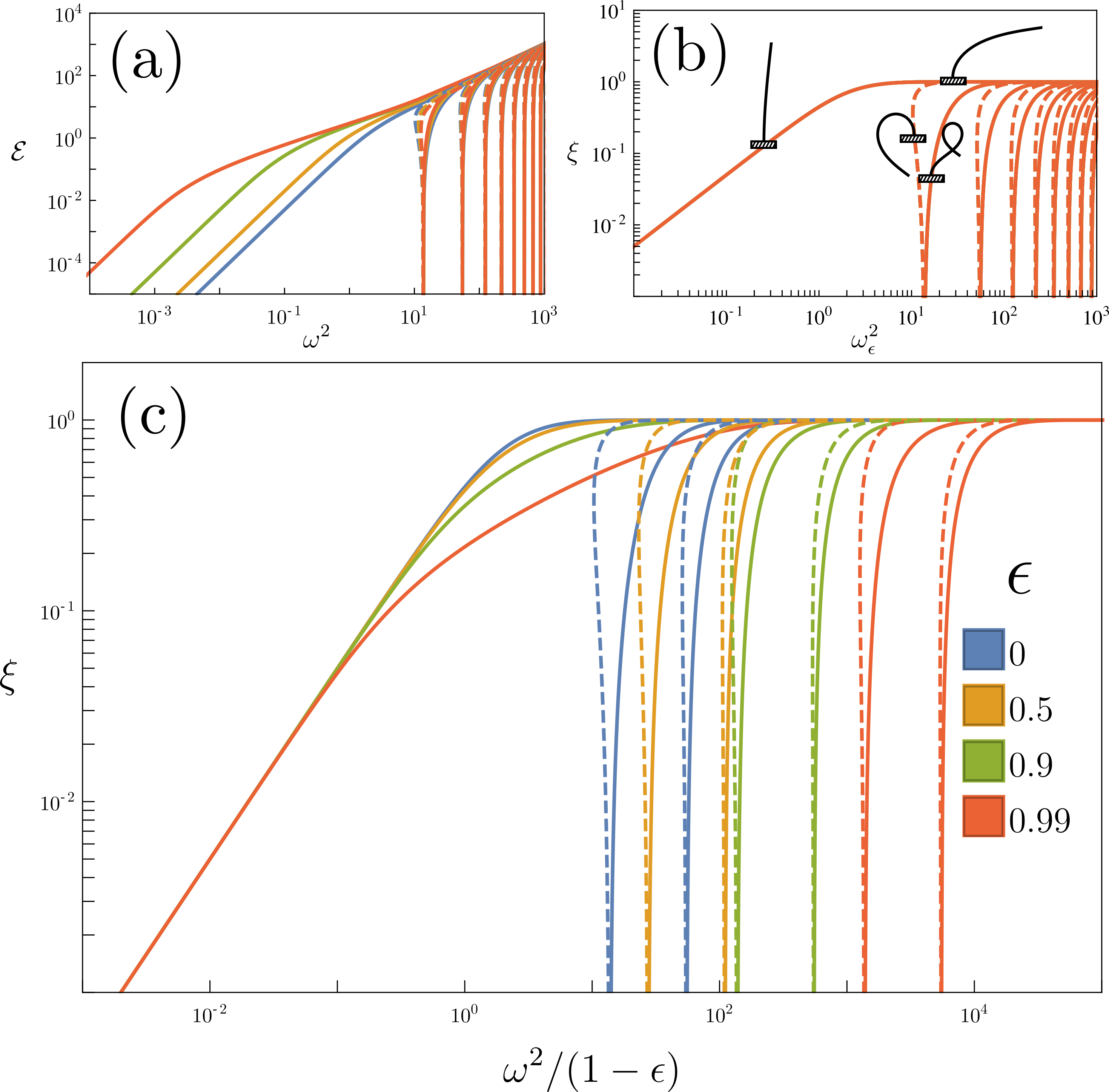}
    \caption{{Plots of solutions of the boundary value problem posed in \cref{sec:formulation} for the case $\theta_0=0$ depending on the two parameters, $\omega^2$ and $\epsilon$. (a) Uncollapsed dependence of the energy (Hamiltonian) $\mathcal{E}$  on the two input parameters. (b) Fully collapsed solution space in terms of the similarity variables $\xi=\mathcal{E}/\omega_\epsilon^2$ vs.\ $\omega_\epsilon^2$, with representative hair profiles. Here, solid and dashed lines indicate positive and negative base curvature, respectively.  Note, $\omega_\epsilon^2$ is a  quantity that depends transcendentally on $\omega^2$ and $\epsilon$.  {For weak and strong forcing we see the predicted scalings of 
 $\xi \approx \omega_\epsilon^2/2$ and $\xi\rightarrow 1$, respectively, with  the crossover occurring near ${\omega_\epsilon^2}/{2} \approx 1$.} (c) Partial collapse of the solution space is seen using the abscissa $\omega^2/(1-\epsilon)$, showing physically realizable branches with an explicit function of the input parameters. To avoid clutter, only the first three branches (and their negative curvature counterparts) are plotted in this panel.}}
\end{figure*}

For the hairs,  choices of $n\neq 0$ and/or  negative curvature branches correspond to ``twirling" profiles (see \cref{fig:solution}). These orbits are not physically realizable for simple shear flow experiments for either of two reasons:
\begin{itemize}
\item Hair profiles intersect the surface they are mounted on (or also themselves). This is possible because the model does not consider hair-surface interactions.
\item The assumption that shear stress is concentrated at the hair tip breaks down because the hair-tip is no longer the portion exposed to shear flow.
\end{itemize}
These solutions are an important consideration nevertheless because numerical algorithms can be susceptible to converging to them. 

All accessible solutions for a discrete list of $\epsilon$ values and a range of $\omega^2$ are plotted in \cref{fig:solution}. In panel (a) we plot the energy $\mathcal{E}$ (a measure of $\theta_0'$) vs.\ $\omega^2$ for the values of $\epsilon$ color coded in panel (c).  The blue curve corresponds to $\epsilon=0$, the case where self-consistency vanishes, while the orange curve shows the distortion caused as   $\epsilon$ approaches unity.  This plot makes it clear that the pendulum analogy alone is insufficient to capture predictions of the basic model.  Panel (b) shows that the solutions of the self-consistent boundary value problem are completely collapsed when the similarity variable $\xi$ is used instead of  $\mathcal{E}$.  In this  plot of $\xi$ vs.\ $\omega_\epsilon^2$  there is only a single curve.  The black lines on this plot  depict  representative hair profiles:  for small $\omega_\epsilon^2$ the hair only slightly bends while there is a scaling change for  $\omega_\epsilon^2 \gtrsim 1$ as the hair bends significantly.  In addition,  for larger $\omega_\epsilon^2$ we obtain the  twirling profiles  where the solid and dashed lines of panel (b)  indicate positive and negative base curvature, respectively. Panel (c) shows that the physically realizable branch can be partially collapsed by plotting $\xi$ vs.\  $\omega^2/(1-\epsilon)$. In the case of the physically realizable solutions, the dependence on $\epsilon$ is most apparent for small forcing where the hair height is maximal.   {For this case (e.g.\ small  imposed fluid velocity $v<<1$) $h_1\rightarrow 1$, i.e., the hair is nearly vertical with $\theta\approx 0$.  Thus from \eqref{intOrig},  $\theta_0'\approx \omega_\epsilon^2$, which with \eqref{E0} gives
\begin{equation}
\xi =\frac{\theta'^2}{2\omega_\epsilon^2} + \sin{\theta}\approx \frac{\theta'^2}{2\omega_\epsilon^2}\approx \frac{\omega_\epsilon^2}{2}\,.  
\end{equation}
This explains the linear dependence and slope observed in panel (b) of  \cref{fig:solution} for small $\omega_\epsilon^2$.  For large forcing where   $\omega^2\rightarrow\infty$, the height of the hair asymptotically approaches zero, i.e., $\theta_1 \approx \pi/2$  and 
 $\xi \approx \sin{\theta_1}\approx 1$, which explains the asymptote of panel (b) of  \cref{fig:solution}.  In this limit
$\omega_\epsilon^2\rightarrow\omega^2$ and the $\epsilon$-dependence vanishes.  Finally, one expects the crossover between weak and strong forcing behavior to occur near ${\omega_\epsilon^2}/{2} \approx 1$, and indeed this is the case. } 

\subsection{Angled hairs: \texorpdfstring{$\theta_0\neq 0$}{t0!=0}}
\label{sec:angHairs}

\begin{figure}
    \centering
    \includegraphics[scale=1.25]{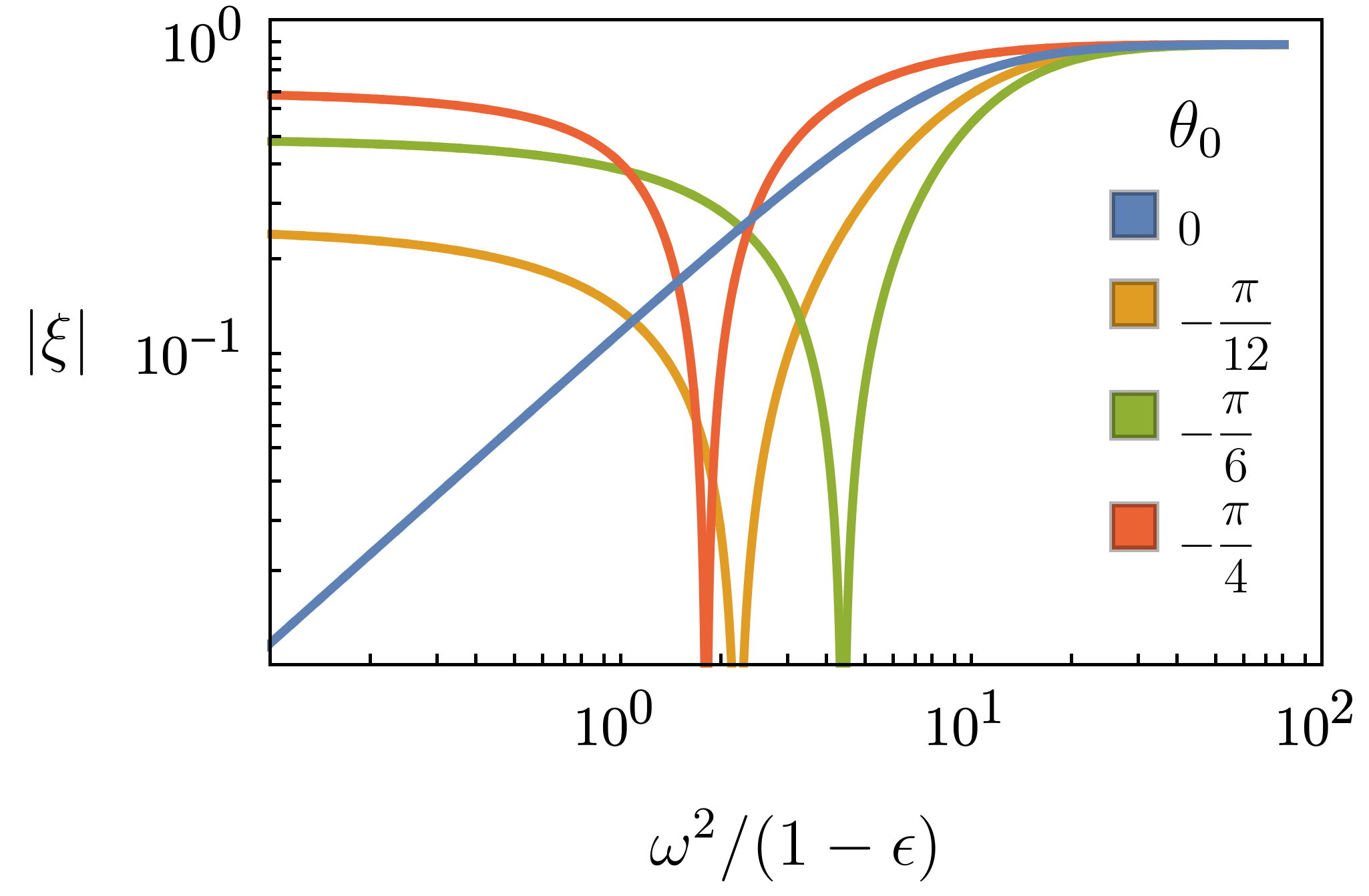}
    \caption{$\xi$ vs $\omega_\epsilon$ plotted using \cref{solution} for different values of $\theta_0$. Hairs with a negative base-angle have negative energy at low $\omega$, and transition to positive energy as $\omega$ increases. Figure created with $\epsilon=0.61$}
    \label{fig:theta0s}
\end{figure}

 Next, we plot $\xi$ vs.\  $\omega_\epsilon$ for different values of $\theta_0$ in \cref{fig:theta0s}. For negative $\theta_0$, shear flow is against the grain. As the forcing increases, hairs reorient to align with the fluid velocity until $\theta_1=0$ which corresponds to $\xi=0$.  Further increasing the forcing parameter brings the system into the flow alignment regime, scaling the same for all $\theta_0$.

On the other hand, increasing $\theta_0$ results in flow with the grain. Flow alignment can be achieved with a smaller forcing parameter (compared to $\theta_0=0$) and the dependence of $\xi$ on $\omega^2/(1-\epsilon)$ approaches a horizontal line.


\section{Hair profiles,  discussion, and comparisons}
\label{sec:hairProfiles}

Recall from \cref{sec:formulation}, the unit tangent is given by  ${\mathbf{R}'}=\hat{\mathbf{T}}$ which implies $x'(\sigma)=\sin\theta(\sigma)$ and $z'(\sigma)= \cos\theta(\sigma)$.  Thus, given our solutions of \cref{sec:solution} for $\theta(\sigma)$, we can plot $z$ vs.\ $x$ for the hair profiles.  In this section we compare hair profiles obtained by our analytic solutions with those obtained by direct numerical integration. A standard numerical method for nonlinear boundary value problems is to use a shooting code, whereby initial values are incremented until the desired boundary value is obtained.   In  \cite{alvarado_nonlinear_2017}  such a shooting code with a standard ordinary differential equation algorithm was used to integrate the pendulum equations of \eqref{TheEqn2}, with an adaptation allowing for the $\theta$-dependence in $\omega_\epsilon$.  Another approach is to make a central difference approximation to the  second derivative of \eqref{TheEqn2}, representing $\theta$ along the centerline of the hair by  a mesh of $N$ segments with values $\theta_i$ ($i=1,2,...,N$). This gives a sequence of  algebraic equations with the boundary conditions built into the first and last equation.  Coupling of the equations  is provided by both the differencing and  the self-consistency through $\omega_\epsilon$.  An example of this procedure is given in \cite{gazzola_forward_2018}, where  the more complicated problem of a filament subject to three dimensional dynamical behavior is solved by discretizing in both space and time.  Associated with this method is a   root finding problem, which for the time-independent case involves solving $N$ equations for each mesh value $\theta_i$. Because there isn't  a concise description of how each of $\theta_i$ asymptotically scales with the forcing parameter, $\omega$, convergence to physical solutions is not always guaranteed.

\begin{figure*}
    \centering
    \includegraphics[scale=1]{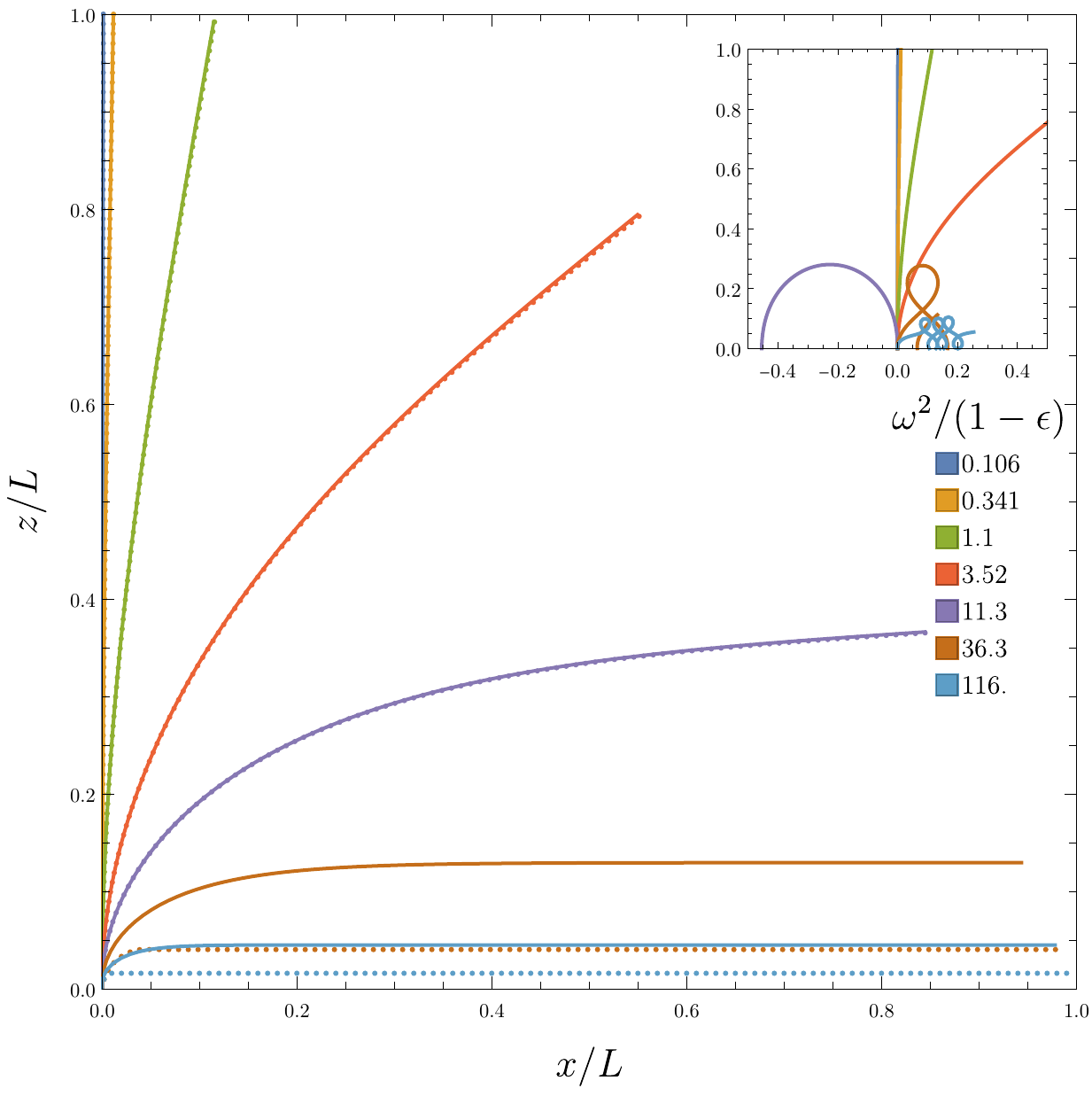}
    \caption{Comparison of  numerical and analytic solutions for a variety of $\mathcal{E}$. Solid and dashed curves indicate numerical and analytic solutions, respectively. A discretization method was used for the numerical routine with an initial estimate of $\theta_i=0$ for the outer figure and $\theta_i=-0.1$ for the inset. For large $\omega$, the numerical solution requires many mesh segments in order to fully resolve the curvature at the base of the hair.}
    \label{fig:profiles}
\end{figure*}

In \cref{fig:profiles} a set of profiles is shown,  comparing our analytic  solution with numerical solutions obtained by  using the mesh discretization described above. At low forcing, both approaches converge to the same, physical solution. At high forcing when the hair becomes more streamlined, the numerical solution requires a larger number of mesh segments in order to fully resolve the high curvature section at the base of the hair and so it does not fully agree with the analytic solution.

%
%
%



Even though there is a root finding problem associated with our analytic solution, it is a single equation (compared to $N$ for the numerical approach). Because of this, and the fact that we know how $\mathcal{E}$ scales in both deformed and undeformed regimes, our analytic method is both simpler to implement and faster to computed  than numerical approaches.

We have observed that the analytic solution is about two orders of magnitude faster ($0.005$\,s vs.\  $0.5$\,s for a finite-difference simulation with a mesh-size of 80) than the numerical procedure. There is not much difference in obtaining a single solution using either approach in terms of speed. However, problems that involve solving \eqref{hairdiffeq} iteratively can benefit significantly from the analytic approach. For example, optimization of the system's rectification properties and solving weakly time-dependent problems ($\omega_\epsilon\rightarrow \omega_\epsilon(t)$) are potentially computationally expensive tasks. 

Given a shear stress, what are the profiles of bed of hairs, which can be dense yet noninteracting? Our solution presented in \eqref{solution} provides an answer to this question. The inverse problem, where the profiles are used to infer the shear stress, is utilized in a recently developed imaging technique. In \cite{liu_measurements_2019,brucker_evidence_2015}, a bed of flexible micropillars is used to detect near-wall shear stress and velocity fields in turbulent flow. The pillars act as wave guides allowing the tip deflection to be measured when illuminated from below. Our analytic method could be used to derive simple expressions for tip-deflection, which can be utilized in the linear, low deformation regime. Greater flow-detection sensitivity can be achieved by increasing the flexibility of the pillars and operating them in the nonlinear regime \cite{brucker_evidence_2015}.

Lastly, we argue that our analytic solution can be used as a basis for understanding  problems where the fluid flow has a slow time dependence. In this regime, a hair cycles through its steady-state profiles, and fluid flows within the hair bed can be neglected.


\section{Summary}
\label{sec:conclusion}
In this work, we obtained a solution to a differential equation describing the profile of a hair bed immersed in shear flow. This problem differs from previous treatments of cantilevered rods in that the forcing parameter has functional dependence on the dependent variable, $\theta(\sigma)$. As a result, the spectrum of permissible $\omega_\epsilon$ at fixed $\xi$ becomes continuous in addition to being discrete. As interesting as they are, many of these solutions are not physically realizable and an advantage of our analytic work is that we can select the desired branch. To contrast this, shooting codes and other numerical approaches cannot be guaranteed to converge to this class of solution.

We then compare the analytic solution to a central difference based numerical scheme that performs reasonably well for the range of loading tested, but can encounter a convergence issue when the curvature at the base is large. 

Future work could explore an adiabatic extension of this model to describe time-dependent channel flows.


\section*{Acknowledgment}
\noindent   PJM was supported by U.S. Dept.\ of Energy Contract \# DE-FG05-80ET-53088.

\appendix

\section{Calculation of \texorpdfstring{$h(\sigma)$}{h(s)}}
\label{appendix}

We wish to calculate $h(\sigma)$ of \eqref{hs}. To this end, let 
\begin{equation}
\label{nonSol2}
    \theta(\sigma)=2\left[\mathrm{am}\left(u\right)- {\pi}/{4}\right]
\quad \mathrm{with}\quad
u:= \frac{\omega_\epsilon}{k}\,\sigma  +F\left(\frac{\pi}{4}+\frac{\theta_0}{2} \right)\,.
\end{equation}
Using elementary trigonometry identities we obtain
\begin{eqnarray}
\label{nonSol3}
 && \cos  \theta(\sigma)=\cos \Big[2\, \big(\mathrm{am}(u)-\pi/4\big)\Big]= 1-2 \sin^2[\mathrm{am}(u)-\pi/4] \,,
\\
&&\sin\big(\mathrm{am}(u)-\pi/4\big) =  \frac{\sqrt{2}}{2}\big( \mathrm{sn}(u) -  \mathrm{cn}(u)\big)\,,
\end{eqnarray}
with the identities sn$(u)=\sin\big(\mathrm{am}(u)\big)$ and cn$(u)=\cos\big(\mathrm{am}(u)\big)$.
Thus, 
\begin{equation}
2\sin^2\big(\mathrm{am}(u)-\pi/4\big) = \big( \mathrm{sn}(u) -  \mathrm{cn}(u)\big)^2
=1- 2\mathrm{cn}(u)\mathrm{sn}(u)\,,
\end{equation}
using $ \mathrm{sn}^2(u)+  \mathrm{cn}^2(u)=1$.  So $1-2\sin^2(u)=2\mathrm{cn}(u)\mathrm{sn}(u)$, 
which with
\begin{equation}
\mathrm{cn}(u)\mathrm{sn}(u)= -\frac{1}{k^2} \frac{d}{du} \mathrm{dn}(u)
\end{equation}
we obtain
\begin{equation}
\int_0^\sigma \cos\theta(\sigma')\, d\sigma'=- \frac{k}{\omega_\epsilon}\, \frac{2}{k^2} \int_{u_0}^u  \frac{d}{du'} \mathrm{dn}(u')\, du'= \frac{2}{k\omega_\epsilon}\big(\mathrm{dn}(u_0)- \mathrm{dn}(u)\big)
\end{equation}
using $d\sigma =  {k}\,du/{\omega_\epsilon}$, where $u_0=  F\left( {\pi}/{4}+ {\theta_0}/{2}\right)$. Finally we use 
\begin{equation}
\mathrm{dn}\big(F(\phi | k^2)\big)= \sqrt{1-k^2\sin^2\phi}
\end{equation}
to obtain the result of \eqref{hs}.





\bibliographystyle{apsrev}

\bibliography{ref}


\end{document}